\documentclass[a4paper,10pt]{article}
\usepackage[utf8]{inputenc}
\usepackage[T1]{fontenc}
\usepackage{amsmath}
\usepackage[symbol]{footmisc}
\usepackage{amsfonts}
\usepackage{setspace}
\spacing{1}
\usepackage{amssymb}
\usepackage{graphicx}
\usepackage[explicit]{titlesec}
\titlespacing\section{0pt}{10pt plus 4pt minus 2pt}{2pt plus 1.5pt minus 1.5pt}
\usepackage{float}
\usepackage{lipsum}
\usepackage{multicol, blindtext, caption}
\usepackage{xcolor}
\usepackage[font=footnotesize,labelfont=bf, skip=10pt]{caption}
\definecolor{azul}{rgb}{0, 0, 0}
\usepackage[left=1.5cm, right=1.5cm, top=2.0cm, bottom=2.5cm]{geometry}
\usepackage[hidelinks]{hyperref}
\author{Ignacio Figueruelo Campanero}
\begin{document}
	

	\begin{center}
		{\Large \textbf{Apparent color and Raman vibrational modes of the unconventional
            superconductor Bi$_2$Sr$_2$CaCu$_2$O$_{8+\delta}$ exfoliated flakes}}\\
		\vspace{5mm}
		{\large Ignacio Figueruelo-Campanero$^{1,2}$\footnote[1]{ignacio.figueruelo@imdea.org}, Adolfo del Campo$^3$, Gladys Nieva$^4$, Elvira M. González$^{1,2}$, Aida Serrano$^3$ and Mariela Menghini$^{1}$\footnote[1]{mariela.menghini@imdea.org}  }\\
		\vspace{3mm}
		$^1$\textit{IMDEA Nanociencia, Cantoblanco 28049, Madrid, Spain}\\
		$^2$\textit{Facultad Ciencias Físicas, Universidad Complutense de Madrid, 28040 Madrid, Spain }\\
        $^3$\textit{Departamento de Electrocerámica, Instituto de Cerámica \textit{y} Vidrio (CSIC), 28049 Madrid, Spain}\\
        $^4$\textit{Laboratorio de Bajas Temperaturas, Centro Atómico Bariloche, 8400 S. C. Bariloche, Argentina}\\
	\end{center}



Studying and controlling the properties of individual exfoliated materials is one of the first steps towards the fabrication of complex van der Waals systems. However, prolonged exposure to ambient conditions can affect the properties of very thin exfoliated materials altering their physical properties. For this reason, it is imperative to employ versatile characterization strategies compatible with reduced ambient exposure times. In this work, we demonstrate that optical microscopy and Raman spectroscopy are quick and non-invasive techniques to study flakes of the high-temperature superconductor Bi$_2$Sr$_2$CaCu$_2$O$_{8+\delta}$ (BSCCO-2212). 
The apparent color of BSCCO-2212 exfoliated flakes on SiO$_2$/Si has been studied allowing a rough and fast identification of the number of layers. Moreover, we find that thin flakes have a refractive index of around 1.7 in the visible range and 0.5 for the absorption coefficient near the maximum at 550 nm. We determine the optimal combination of illumination wavelength and substrate properties for the identification of different numbers of unit cells of BSCCO-2212. In addition, we report the hardening of the characteristic Raman modes at 116 cm$^{-1}$ and 460 cm$^{-1}$ as flake thickness decreases, possibly due to strain in the BiO and CuO$_2$ planes, respectively. Moreover, the evolution of the Raman modes establishes a second approach to determine the thickness of BSCCO-2212 thin flakes. As BSCCO-2212 is a challenging material to be due to its sensitivity to ambient conditions deriving in an insulating state, the present work provides a guide for the fabrication and characterization of complex van der Waals systems paving the way for studying heterostructures based on unconventional superconductors in the 2D limit.\\\\
		\vspace{2mm}
		\underline{\textbf{Keywords:}} \hspace{2mm} \textit{BSCCO-2212, mechanical exfoliation, refractive index, Raman spectroscopy}


	\begin{multicols}{2}
		\section*{I. Introduction} \label{Intro}

Since the isolation of graphene from bulk graphite crystals \cite{Novoselov_2004}, van der Waals (vdW) materials have attracted a lot of interest due to their different properties compared to their bulk counterparts \cite{Novoselov_2016, Geim_2013}. Some of these materials, also termed as 2D materials, belong to the broader class of quantum materials \cite{2020, Basov_2017} and can host superconductivity \cite{Khestanova_2018}, topological phases \cite{Kou_2017}, correlated electronic states \cite{Cao_2018} and magnetism \cite{Gibertini_2019} among other phenomena. Due to their very attractive properties, 2D materials are a great playground to study quantum phenomena. The possibility of stacking different types of materials, in a relatively easy way, allows the creation of heterostructures with on-demand or novel properties. High temperature superconductors (HTSC) shine for their complex and rich phase diagram as a function of doping \cite{Rybicki_2016}. Specifically, Bi$_2$Sr$_2$CaCu$_2$O$_{8+\delta}$ (BSCCO-2212) is one of the most studied HTSC \cite{Zhu_2021, Sterpetti_2017} and one of the few that exhibits vdW bonds bridging HTSC and 2D materials. One or half unit cell (UC) of BSCCO-2212 has claimed attention as it stands as a perfect platform to study non-conventional 2D superconductivity \cite{Yu_2019,Zhao_2023}. Besides, it has been demonstrated the possibility to fabricate single photon detectors \cite{Merino_2023} and nanomechanical oscillators \cite{Sahu_2019} based on exfoliated BSCCO-2212. Although it is a promising material, only a few works \cite{Poccia_2020, Lupascu_2012} have studied BSCCO-2212 exfoliated flakes in depth due to the difficulty in its manipulation. Exposure of the flakes to ambient conditions can change the concentration of oxygen content, $\delta$, destroying the metallic-superconducting behavior and becoming insulating \cite{Yu_2019,Zhao_2020,Sandilands_2014,Saito_2005}. In this sense, BSCCO-2212 must be carefully prepared in controlled inert atmospheres such as Ar or N$_2$ glove-boxes, even using liquid nitrogen cooled stages \cite{Zhao_2023}, making especially difficult to perform fundamental studies and to integrate it with other commonly used 2D materials.\\
Optical microscopy and Raman spectroscopy are two of the most common tools used for studying 2D materials. On one hand, optical characterization is a direct non-destructive technique compatible with glove-box set-ups. The different colors that 2D materials in the form of flakes possess when transferred to different substrates, called apparent color \cite{Puebla_2022}, gives information about optical properties and the thickness of the flakes \cite{Zhang_2021,Zhao_2020B}.  The quantification of the flake colors by means of optical contrast (OC) \cite{Blake_2007} when illuminating with monochromatic light, is a powerful characterization technique allowing rapid discrimination of flakes with different thicknesses through a simple optical inspection. Moreover, the OC can be modeled by a Fresnel multilayered system to extract optical properties such as the refractive index of the material. On the other hand, micro-Raman spectroscopy is a non-invasive and fast characterization technique very widely used to study 2D materials. This technique gives very useful information about structural and electronic properties of materials and has been used to check composition, identify unwanted by-products, chemical modifications or even structural damages during exfoliation or transfer of vdW materials \cite{Castellanos_Gomez_2013}. Furthermore, Raman spectroscopy can be used to determine the thickness of exfoliated materials, since vibrational modes can be sensitive to the number of layers as reported for 2D materials such as graphene \cite{Koh_2010} or MoS$_2$ \cite{Molina_S_nchez_2011}. However, for some materials and depending on the experimental conditions \cite{Castellanos_Gomez_2013,Wang_2023}, the Raman laser power can induce damage on very thin flakes under prolonged exposure time, which must be considered for a proper preliminary characterization.\\
In this work, we report an experimental study on the properties of thin BSCCO-2212 flakes by combining optical microscopy and micro-Raman spectroscopy. We show how apparent color is a reliable tool to identify a wide range of thicknesses for BSCCO-2212 flakes down to few layers. From the OC, the refractive index is obtained and we determine the best conditions to identify 1 UC of BSCCO-2212 on top of SiO$_2$/Si substrates. Finally, by means of a micro-Raman spectroscopy we study the vibrational modes of thin flakes, finding a Raman hardening for the most prominent vibrational modes as a function of thickness.
\section*{II. Experimental methods} \label{exp}
BSCCO-2212 flakes were exfoliated in a N$_2$ glove-box from optimally doped single crystals grown by the self- flux method \cite{Correa_2001}. BSCCO-2212 has a layered perovskite structure formed by stacked planes in the c-direction as represented in Figure 1a. The unit cell of this compound is 3.07 nm high and is formed by repeating another half unit cell on top of the one presented in Fig. 1a shifted by (1/2,0,0). vdW bonds are present between the stacked BiO planes of the material and thus mechanical exfoliation permits the obtention of ultra-thin BSCCO-2212 flakes. In our case, Nitto tape (SPV 224) was used to cleave the single crystal. The substrates used for the characterization were Si (Siltronix) p-doped (100) with a top layer of SiO$_2$ 290 nm thick. The stamping was done by gently pressing the tape onto the wafer for a few minutes and slowly releasing it. In the present work, a total of 40 different flakes with thickness up to 150 nm were investigated.\\
A commercial Park XE7 atomic force microscopy (AFM) system in tapping mode was employed to determine the thickness of the exfoliated BSCCO-2212 flakes. The AFM probes were commercial Si cantilever NANOSENSORS AC160TS. All presented thicknesses have a standard deviation of maximum 15$\%$ of their mean values.
Optical images of the transferred flakes were obtained using a Canon EOS 1300D camera attached to a Nikon Eclipse Ci optical microscope in reflection mode with a x40 objective (numerical aperture, NA=0.65). Monochromatic illumination was obtained using band-pass filters of 450 nm, 490 nm, 520 nm, 532 nm, 550 nm, 580 nm, 610 nm and 650 nm from ThorLabs with a full width high maximum (FWHM) of 10 nm except for the 532 nm with 5 nm FWHM.\\
For Raman measurements we used a confocal microscope WITec ALPHA 300 RA, equipped with a Nd:YAG p-polarized laser at 532 nm excitation. Measurements were performed using x100 objective (NA=0.95), using an 1800 g/mm grating and a laser output power of 3 mW to minimize overheating effects and flake degradation. Raman data were analyzed by WITec Plus Software 2.08. 

\section*{III. Results and Discussion}
Fig. 1b shows the apparent color under an optical microscope examination of BSCCO-2212 exfoliated flakes on a SiO$_2$/Si substrate. By measuring the height profile by AFM of the different flakes a color-thickness chart is built ranging from 0 to 150 nm ($\sim $ 50 UC), as shown at the bottom of Fig. 1b. Apparent color is the result of the combination between absorption and interference of light. Ultra-thin flakes present blue-purple tonalities, very similar to the substrate. In this limit, flakes barely attenuate the incident light, also the interference is negligible and thus the resulting colors come mainly from the absorption of the Si and the interference in the SiO$_2$ layer, making different materials to look similar when exfoliated down to very few layers on similar substrates \cite{Puebla_2022,M_ller_2015}. As flake thickness increases, tonalities shift to green - yellow for around 30 – 80 nm, and to orange - red for 80-130 nm, in agreement with Puebla \textit{et al.} \cite{Puebla_2022}. For thick enough flakes the interference of light within the flake and the absorption start to be significant and consequently the contribution from the flake starts to dominate the apparent color.
As reported in other works \cite{Puebla_2022, Papadopoulos_2018}, the apparent color seems to be a periodic function of the thickness, for example for a thickness of 150 nm a purple tonality is found again but much darker than the one present in very thin layers. As thickness increases beyond 150 nm similar colors as the ones reported but in darker tonalities appear until brown tones, where there is no contribution from the SiO$_2$/Si, and the overall reflected light comes from the BSCCO-2212. These thick flakes do not present much interest as their properties are almost the same as those of the bulk material and they present difficulties when transferring onto substrates due to the formation of air bubbles, wrinkles and inhomogeneities.\\ 
The RGB components of selected flakes were extracted from the color chart for their respective thickness and the results are presented in Fig. 1c. For the red channel, the values do not show strong dependence on the thickness, as it ranges from 125 arb.u. to 225 arb.u. This component dominates for thicknesses higher than 90 nm (red tonalities) and is minimum from 20 to 40 nm (blue tonalities). For the green channel, it is notorious how the component increases as thickness increases in the range between 0 to 50 nm. Between 50 and 80 nm, it remains constant (flakes of green tonalities) reaching a maximum value at 80 nm. From there, the green component decreases and seems to become constant at around 130 nm. Finally, for the blue channel we observe that it predominates over the two other components for ultra-thin flakes (up to 30 nm). No dependence can be seen for very thin flakes in the range 0-60 nm (blue to greenish tonalities), and then it has a minimum of 60 arb.u. at 95 nm where it starts to slowly recover for larger thickness (maroon-purple tonalities).From the constructed graph we can delimit the different thickness regions according to their RGB component ratio. These thickness-dependent regions can be used in other experimental setups to have a rough estimation of the flake thickness under test as the relative contribution of RGB channels (e.g. R>G>B) will be maintained even if the absolute values are different as these will depend on the illumination source and camera settings).\\
As mentioned above, when exposed to air, BSCCO-2212 flakes lose oxygen dopant content which results in an abrupt change in resistivity \cite{Yu_2019} and consequently in a metal to insulator transition. We have characterized the apparent colors of flakes inside and outside a N$_2$ glove-box to check if there were changes in tonalities due to ambient exposure and thus to the optical properties of the material. We found the same tonalities after a few days for all the flakes outside the glove-box. This indicates that optical properties are not fully defined by doping. The difference between metallic-superconductor (optimally doped) and insulating (underdoped) samples is around 0.16 holes per Cu atom \cite{Presland_1991}. We conclude that the mean refractive index of the whole sample remains almost constant with time and apparent color is not influenced by the possible oxygen variation due to degradation.\\
\begin{figure*}
    \centering
    \includegraphics[width=0.8\textwidth]{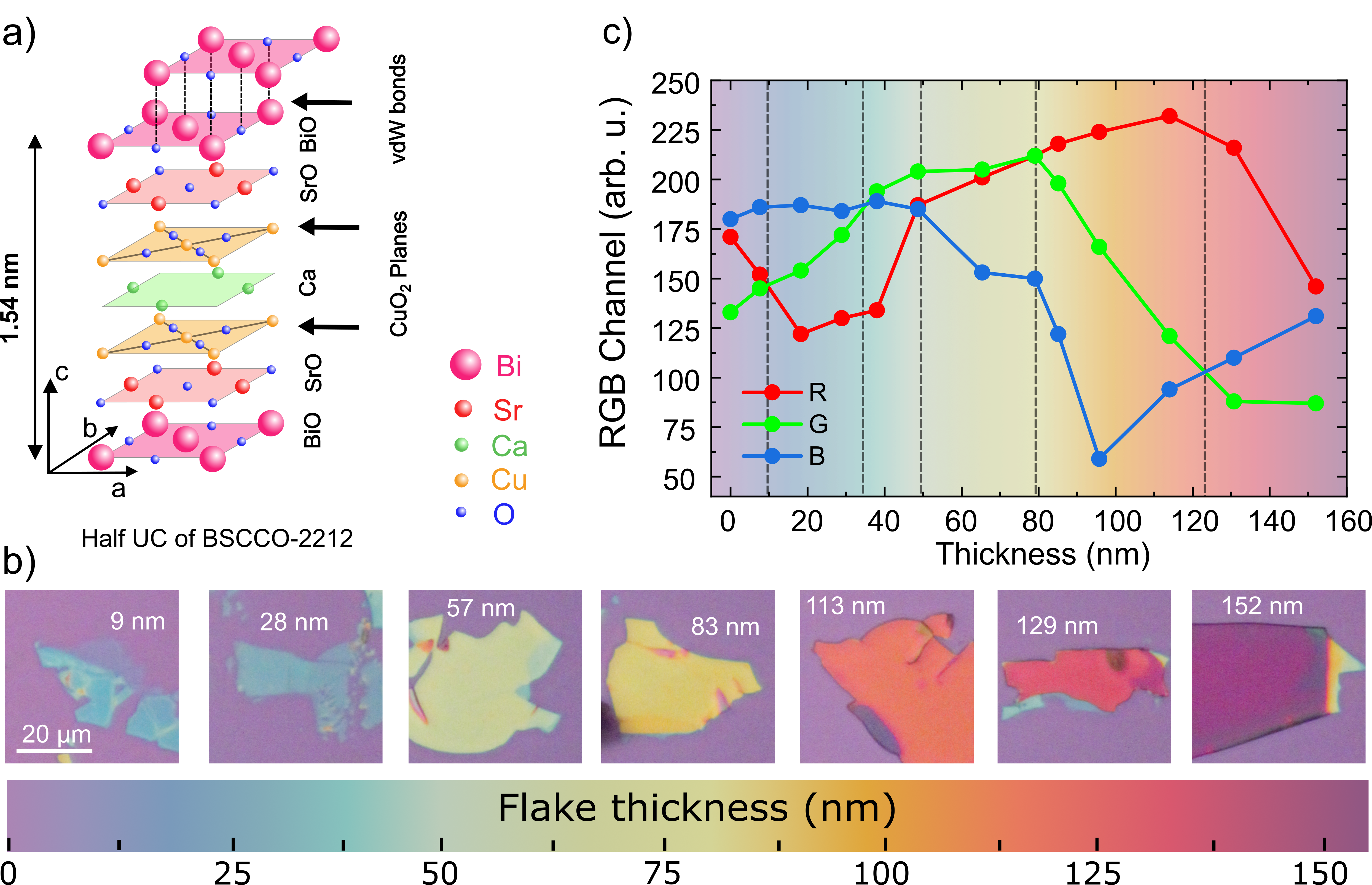}
    \caption{Apparent color of BSCCO-2212 exfoliated flakes. a) Half unit cell of BSCCO-2212. b) Top: Optical images in reflection mode of different flakes with thicknesses ranging from 9 nm ($\sim$3 UC) to 150 nm ($\sim$50 UC) transferred on top of a Si substrate with a 290 nm capping of SiO$_2$ . Bottom: Color scale showing the apparent color - thickness chart of BSCCO 2212 exfoliated flakes. c) Extracted RGB composition vs thickness of selected flakes. Dotted lines represent the thicknesses where the RGB components change their relative order so it can be used for rough identification.}
    \label{Fig1}
\end{figure*}
Although apparent color can be a first approximation tool for thickness identification, it strongly depends on the source of illumination and the settings on the camera. However, by illuminating with monochromatic light of different wavelengths ($\lambda$) a more profound and reproducible study based on the OC can be done to extract information about flake properties \cite{Blake_2007}. The flake OC can be calculated experimentally as:
\begin{equation}
    OC=\frac{I_F-I_S}{I_F+I_S}
\end{equation}
where $I_F$  is the intensity of the illuminated flakes and $I_S$ is the reflected intensity from the substrate near the flake. Careful attention must be taken in order to not overexpose the pixels of the camera as this can give wrong OC values. Experimental values of the OC for the different band pass filters are represented in Figure 2a. For low wavelengths (blue-like) OC is near 0 for an extended range of thickness values of thin flakes. For $\lambda = 450$ nm, this range covers almost from 0 to 50 nm making it very difficult to identify flakes from the substrate. As we increase the illumination wavelength, for very thin flake, the OC starts to be positive up to $\lambda=550$ nm. For $\lambda = 580$ nm the OC develops a minimum near thicknesses between 7-10 nm that starts to shift to larger thickness as the wavelength increases. This makes the wavelength range 610-650 nm an optimal region to identify very thin BSCCO-2212 flakes. Another minimum can be spotted for larger thicknesses (around 80 nm for $\lambda =450$ nm) that also shifts to larger thickness when increasing wavelength. These minima in the OC appear due to a decrease in the light re-emitted by the flakes and are related to destructive interference in the BSCCO-2212/SiO$_2$/Si multilayer system.\\
A flake composed of regions with two different thickness is presented in Fig. 2b as an example of how OC varies for the different illumination conditions. On the one hand, for the 115 nm thick region, the OC is negative for $\lambda = 450$ nm and as wavelength increases it slowly becomes less dark (smaller OC) until $\lambda = 550$ nm where it shows a positive OC that maximizes at $\lambda = 650$ nm. On the other hand, for the 65 nm thick region,the OC is also negative for $\lambda = 450$ nm  but has a positive maximum at around $\lambda = 580 $nm.\\
To explain the observed experimental OC values,the system was modeled applying Fresnel laws to a multilayer composed by air/BSCCO-2212/SiO$_2$/Si, similar to previous works \cite{Castellanos_Gomez_2013, Koh_2010, Bing_2018}. Under perpendicular illumination the reflected intensity from the modelled heterostructure has the form:
\begin{figure*}
    \centering
    \includegraphics[width=0.8\textwidth]{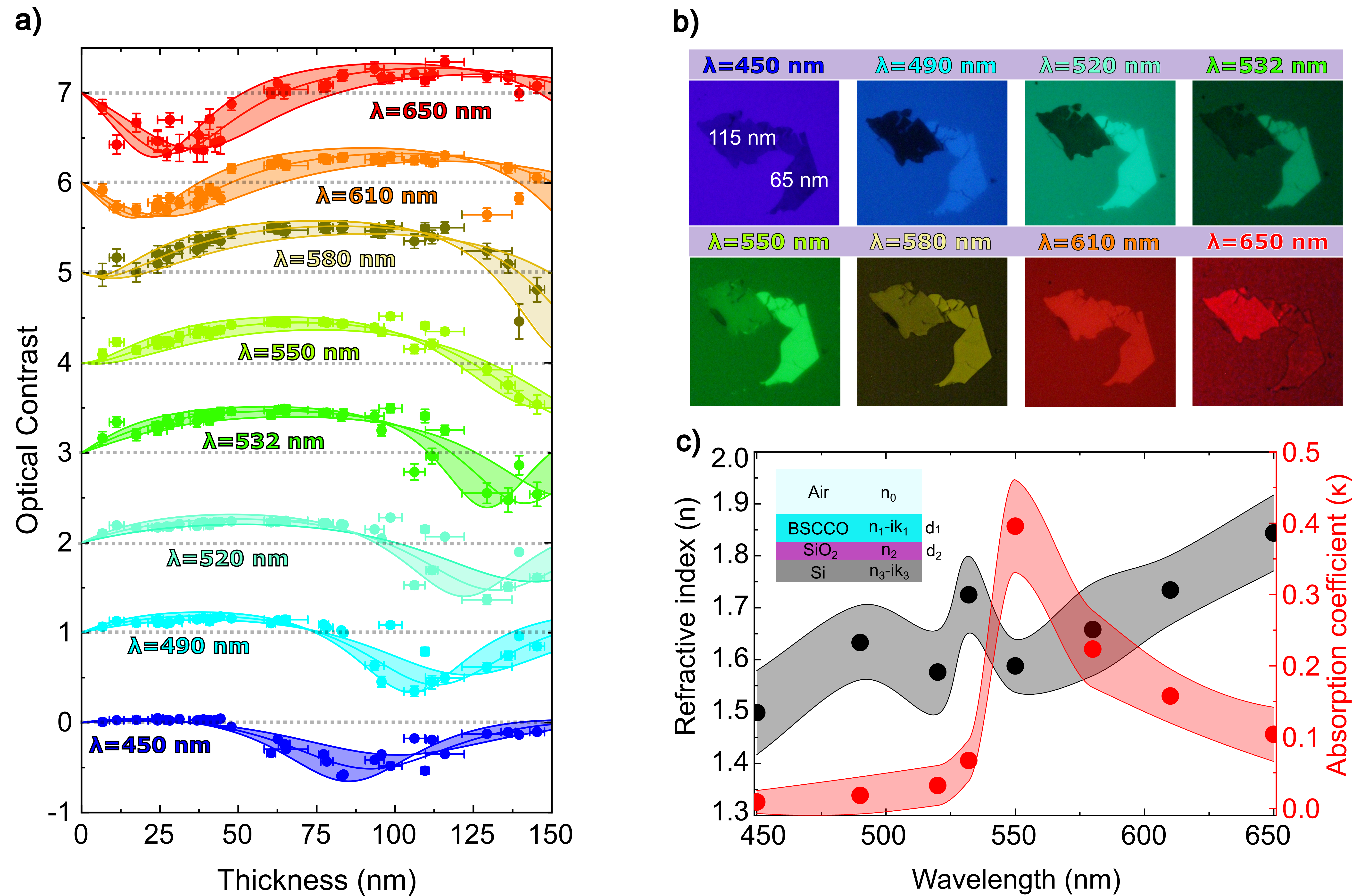}
    \caption{OC characterization of BSCCO 2212 flakes. a) Experimental OC (circles) for different flake thicknesses under monochromatic illumination: 450 nm (blue), 490 nm (cyan), 520 nm (blue-green), 532 nm (green), 550 nm (light green), 580 nm (yellow), 610 nm (orange) and 650 nm (red). Solid lines correspond to theoretical OC fittings calculated using Eqs. (1) and (2) (see text). OC values for wavelengths from 490nm to 650 nm have been shifted vertically for clarity. Shaded regions consider a 10$\%$ deviation in $n$ and $\kappa$ obtained from the fits. b) Two-height BSCCO-2212 flake under different monochromatic illumination using the bandwidth filters. c) Real and imaginary part of the complex refractive index of BSCCO-2212 for the different wavelength illumination, obtained from the OC fittings. Shaded regions represent the error given by the fittings. Inset: Schematics of the multilayer system modeled to fit experimental data in a). }
    \label{Fig2}
\end{figure*}
\end{multicols}

\begin{equation}
    I=\left| \frac{r_{01}e^{i(\phi_1+\phi_2)}+r_{12}e^{-i(\phi_1-\phi_2)}+r_{23}e^{-i(\phi_1+\phi_2)}+r_{01}r_{12}r_{23}e^{i(\phi_1-\phi_2)}}{e^{i(\phi_1+\phi_2)}+r_{01}r_{12}e^{-i(\phi_1-\phi_2)}+r_{01}r_{23}e^{-i(\phi_1+\phi_2)}+r_{12}r_{23}e^{i(\phi_1-\phi_2)}} \right|^2
\end{equation}
\vspace{0.2mm}
\begin{multicols}{2}
where the subindices correspond to the different media: air (0), BSCCO-2212 (1), SiO$_2$ (2) and Si (3). $r_{kl}=\frac{(\tilde{n}_k-\tilde{n}_l)}{(\tilde{n}_k+\tilde{n}_l)}$ are the Fresnel coefficients for the different media and $\phi_k=\frac{2\pi\tilde{n}_k d_k} {\lambda }$ are the phase acquired in medium $\kappa$ with thickness $d_k$. We consider complex refractive indices as $\tilde{n}=n-i\kappa$. The substrate intensity ($I_S$) can be obtained using the same equation replacing the flake for air in medium 1.
We use the formulas (1) and (2) to fit the experimental data, leaving BSCCO-2212 flakes refractive index as a fitting parameter. This approximation assumes no thickness dependence for the refractive index. This is a valid assumption according to reported results on other 2D materials, such as graphene and MoS$_2$, where no significant dependence of refractive index with thickness is observed \cite{Castellanos_Gomez_2013,Koh_2010}. For the SiO$_2$ and Si substrates, we let the refractive and absorption coefficient vary 5$\%$ from tabulated values \cite{Schinke_2015,Gao_2013}. The theoretical OC as a function of thickness for different wavelengths are represented as continuous lines in Fig. 2a. As it can be observed, theoretical results adjust reasonably well to the experimental values. Uncertainty is added to the calculations (shaded colored areas) by varying a factor of $\approx$ 10$\%$ the refractive index obtained from the fittings to consider the possible error of assuming a thickness-independent refractive index for the BSCCO-2212 flakes.
The real and imaginary part of the calculated refractive index in the range from $\lambda = 450$ nm to $\lambda = 650$ nm based on the experimental data of our flakes is displayed in Fig. 2c. For the real part, we observe an overall incremental behavior as the wavelength increases. For the absorption coefficient ($\kappa$), an abrupt increase near 550 nm is observed from almost 0 to 0.4 while it slowly decreases as the wavelength further increases.\\
To our knowledge, there are only a few reports on the optical properties of BSCCO-2212 in the visible range in the literature \cite{Kobayashi_1996,Ismail_2021, Hussein_2019}. For bulk samples, a refractive index for BSCCO-2212 around 1.9 – 2 for the real part and 0.3 – 0.5 for the imaginary part for an illumination wavelength of 488 nm has been reported by polarimeter measurements \cite{Kobayashi_1996}. Ismail \textit{et al.} \cite{Ismail_2021} obtained a refractive index for BSCCO-2212 thin films with a real part around 2 and an imaginary part of 0.2-0.5 depending on the growing conditions. Similar values are obtained by Hussein \textit{et al.} \cite{Hussein_2019} for thin films of BSCCO-2223 with an almost constant real part of the refractive index around 1.8 and an imaginary part around 0.4 – 0.5. None of these works reports the dependency of n and $\kappa$ with wavelength but they observe that the optical properties of the samples change depending on the sample growth method and conditions. In addition, it is worth mentioning that we are considering just one layer of homogeneous BSCCO-2212 for the calculations. A more detailed model should consider the roughness of the flakes as well as a possible spatially heterogeneous distribution of optical constants due to oxygen content variations. In any case, our results for both $n$ and $\kappa$ are similar to reported values for bulk and thin films.\\\\
To provide an easy guide to the exfoliation of BSCCO-2212 flakes, we have also calculated the OC using equations (1) and (2) for different thickness of the SiO$_2$ layer of the substrate under different illumination wavelengths and for different BSCCO-2212 thicknesses, see Figure 3. For the OC calculations, we used the obtained BSCCO-2212 refractive index (see Fig. 2c). We approximated the real part by a quadratic wavelength dependence as the calculations show a monotonous increasing behavior, except for the point corresponding to $\lambda= 532$ nm. For the absorption coefficient, $\kappa$, we used a modified Akima interpolation taking care that the maximum is located close to the observed maximum around 550 nm. Due to the available wavelengths used for this study, we can expect that the actual maximum of $\kappa$ is located in the range between 540 nm and 580 nm, thus our results could be modified shifting the minima in the OC (see Fig. 3) and changing mostly the value of the OC in that region.\\
In Fig. 3a we present the OC mapping covering all visible spectra and different thickness of SiO$_2$ layer for 1 UC of BSCCO-2212. Very small OC can be found when there is no capping layer, demonstrating the importance of the SiO$_2$ layer for optical identification. Also, some other regions present null OC for SiO$_2$ thicknesses between $\approx 70-110$ nm and from 150-300 nm (regions of OC = 0 are indicated in the map by a dotted line). As previously mentioned, we found an absorption peak around 550 nm, and this is reflected in the obtained calculation as the best wavelength to identify 1 UC of BSCCO-2212. \\
For this wavelength we obtain that for 550 nm illumination the best SiO$_2$ layer thickness for identification of 1 UC would be either 80 nm or 260 nm. Fig. 3b and 3c show the OC for thicknesses ranging from 1 to 10 UC (roughly 3 - 30 nm) for the most common commercial substrates used in laboratories with SiO$_2$ layer thickness of 90 nm and 290 nm. For 290 nm SiO$_2$ layer (substrates used in this work) the OC calculated for the 1 UC nearly reaches -0.2, while for thicker flakes the larger obtained values approach 0.4 which agrees with our experimental results. For 90 nm thickness, the OC values are larger than for 290 nm. We conclude that for BSCCO-2212 a SiO$_2$ layer of 90 nm will result in better OC values than 290 nm, improving 1 UC identification. OC maps for different SiO$_2$ layers and for up to 100 nm BSCCO-2212 thick flakes are shown in Figure S3 in the Supporting Information (SI).

\begin{figure}[H]
	\includegraphics[width=\linewidth]{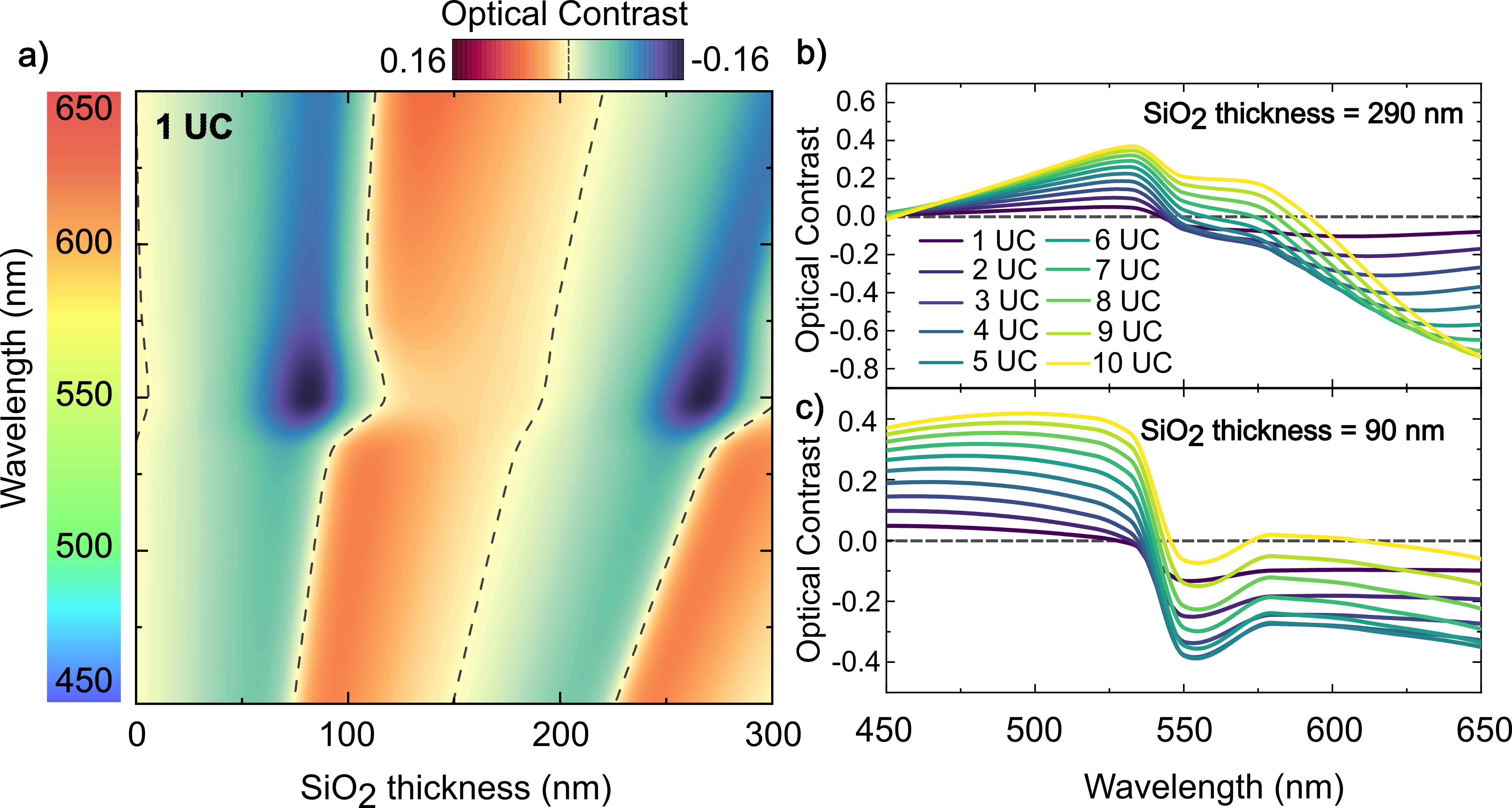}
    \caption{Calculated OC of BSCCO 2212. a) Calculated OC map for 1 UC of BSCCO-2212 for different illumination wavelength and different capping layers of SiO$_2$ derived from Eqs.(1) and (2) and using interpolated values for BSCCO-2212 refractive index (Fig. 2c). OC=0 is marked by dotted lines. OC calculated as a function of illumination wavelength for a fixed SiO$_2$ thickness of b) 290 nm and c) 90 nm for thicknesses between 1 to 10 UC.}
    \label{Fig3}
\end{figure}
\begin{figure*}
    \centering
    \includegraphics[width=1\textwidth]{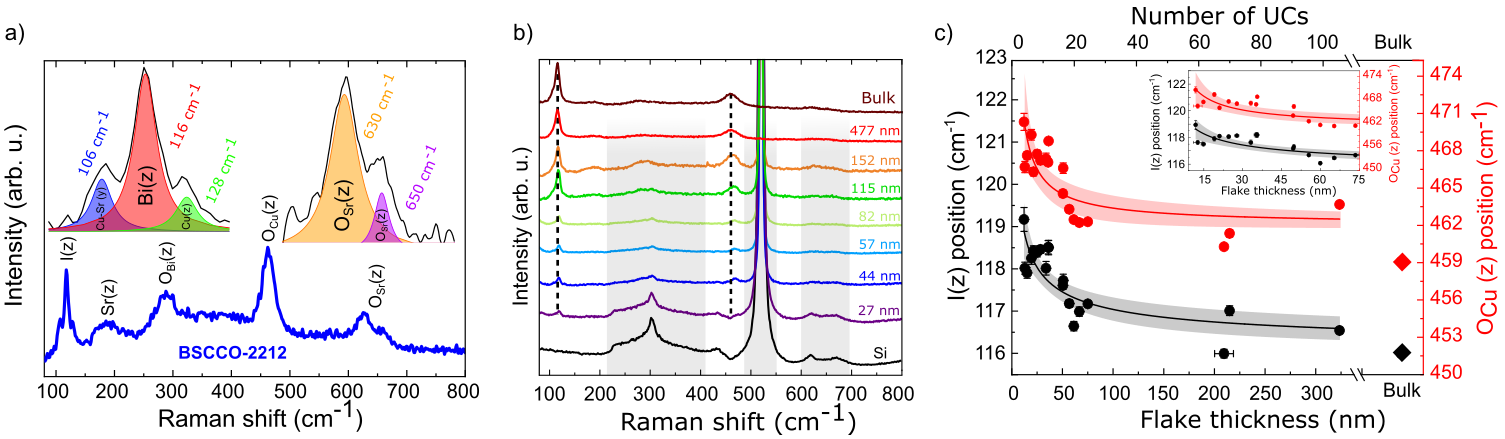}    \caption{Micro-Raman characterization of BSCCO 2212 flakes. a) Raman spectra for BSCCO-2212 bulk sample measured with a 532 nm laser at 3 mW. The Raman mode assignment are done according to Yelpo \textit{et al.} [32]. Inset: Deconvolution of Raman modes at 106-116-128 cm$^{-1}$ and 630-650 cm$^{-1}$. b) Raman spectra for BSCCO-2212 exfoliated flakes with different thicknesses and SiO$_2$/Si substrate Raman spectrum (black). Grey shaded regions indicate areas where the SiO$_2$/Si Raman signal has a contribution in the Raman spectra of BSCCO-2212. c) Evolution of Raman modes at 116 cm$^{-1}$ (black circles) and 460 cm$^{-1}$ (red circles) with thickness. Experimental data are adjusted according to the phenomenological Eq. (3) [60] (black and red lines) with their respective confidence bounds (shaded regions). Inset: Zoomed region from 0 to 75 nm flake thickness. }
    \label{Fig4}
\end{figure*}

In order to evaluate the Raman characteristics of BSCCO-2212 flakes depending on the thickness, a micro-Raman study was carried out on a series of exfoliated flakes of different thickness at room temperature and compared with the spectrum of a BSCCO-2212 single crystal. The results for the single crystal are presented in Figure 4a. BSCCO-2212 has 6 prominent Raman bands between 100 and 700 cm$^{-1}$. There is controversy in the literature about the origin of these bands where different works have reported different mode assignments. A thorough compilation of the different peak assignments for BSCCO-2212 single crystals from different groups together with an \textit{ab-initio} calculation of the phonon frequencies and eigenvectors of the Raman active modes was performed by Yelpo \textit{et al.} \cite{Yelpo_2021}. The first peak for our bulk sample is located around 116 cm$^{-1}$. A closer look to this region of the spectrum (see inserts in Fig. 4a) shows that this band is the convolution of 3 bands centered at 106 cm$^{-1}$, 116 cm$^{-1}$ and 130 cm$^{-1}$. The first band located at 106 cm$^{-1}$ has been assigned experimentally to oscillations in the \textit{y} direction of either Sr \cite{Sugai_1990} or Bi atoms \cite{Klein_1982}, however theoretical calculations \cite{Yelpo_2021} showed that this band could correspond to a Cu-Sr stretching mode along the \textit{y} axis located at 95 cm$^{-1}$. According to the literature, a Raman peak around 116 cm$^{-1}$ is related to stretching of atoms in the \textit{z} direction, but there is discordance in which atoms are involved, as some authors report \textit{z} stretching modes of Bi \cite{Klein_1982, Martin_1995, Cardona_1988} and others of Sr \cite{Sugai_1990,Denisov_1989, Boekholt_1990,Osada_1996}. Although there is discrepancy in the assignment of the peak observed around 116 cm$^{-1}$, both experimental and theoretical works point to vibrations in the \textit{z}-direction. Therefore, we will label this mode as I(\textit{z}) in the following discussion. Finally, for the last band in the convolution, previous works assign vibrations of Cu atoms along the \textit{z} direction to a band located around 128 cm$^{-1}$ \cite{Klein_1982, Martin_1995}. The \textit{ab-initio} calculations presented in \cite{Yelpo_2021} predict a Raman band at 121 cm$^{-1}$ corresponding to a Cu(\textit{z}) mode. The next observed Raman mode in BSCCO-2212 single crystal is located at 185 cm$^{-1}$, and it has been assigned to Sr vibrations along the \textit{z} axis according to experimental \cite{Martin_1995,Cardona_1988} and theoretical works \cite{Yelpo_2021, Kovaleva_2004, Falter_2003}. In the region of 290 cm$^{-1}$ there is another band that could be associated with the convolution of 2 Raman modes separated by around 10 cm$^{-1}$ and possibly due to vibrations along the \textit{z} direction of the O atoms bonded to Cu or Bi atoms \cite{Yelpo_2021,Sugai_1990, Klein_1982, Martin_1995, Cardona_1988, Denisov_1989, Boekholt_1990, Osada_1996, Falter_2003, Chen_1998}. Near 460 cm$^{-1}$ a clear Raman peak is observed and as for the previous mode, there is some discrepancy in the literature on the assignment of this band. According to references \cite{Yelpo_2021, Boekholt_1990} this band can be due to O-Cu vibrations in the \textit{z} direction, while others point to O-Sr(\textit{z})  \cite{Sugai_1990, Klein_1982, Martin_1995, Cardona_1988} or O-Bi(\textit{z}) vibrations \cite{Denisov_1989, Osada_1996}. Finally, between 600 and 700 cm$^{-1}$ there are two other bands located at 630 cm$^{-1}$ and 650 cm$^{-1}$. Following different works \cite{Sugai_1990, Klein_1982, Martin_1995, Cardona_1988, Denisov_1989, Boekholt_1990, Osada_1996, Holiastou_1997}, both bands are assigned to O-Bi(\textit{z}) or O-Sr(\textit{z}), however, according to calculations \cite{Yelpo_2021, Kovaleva_2004, Falter_2003} they should be assigned to apical oxygen (O$_{Sr}$).\\
Raman results obtained for different BSCCO-2212 flakes are displayed in Fig. 4b. For thick enough flakes, all bands can be clearly identified as described previously. However, SiO$_2$/Si substrates exhibit some bands at similar Raman shifts to those of BSCCO-2212, as shown by the gray shaded areas. For very thin flakes the identification of these bands becomes complicated as the Raman intensity from SiO$_2$/Si is stronger than the BSCCO-2212 one. This is the case for thin flakes (see SI) where most of the Si signal covers all the Raman spectra from the BSCCO-2212 flakes. For this reason, we have focused the analysis on samples where the Raman spectra is still discernible from the one of Si, and on bands that do not overlap with those of SiO$_2$/Si bands with the largest intensity, i.e., those at 116 cm$^{-1}$ and 460 cm$^{-1}$ identified as I(\textit{z}) and O-Cu(\textit{z}), respectively.\\
The evolution of Raman spectra with thickness is presented in Fig. 4c. For the most prominent modes at 116 cm$^{-1}$ and 460 cm$^{-1}$ we observe a shift towards higher wavenumbers, i.e., Raman hardening, as the number of layers decreases.This effect is more prominent below 200 nm thickness. The observed shift for the lower band ranges from 116.2 cm$^{-1}$ for bulk up to 119.5 cm$^{-1}$ for very thin flakes ($\Delta$ $\approx$ 3 cm$^{-1}$) while for the higher band, bulk values are around 460 cm$^{-1}$ while for thin flakes it moves up to around 470 cm$^{-1}$ ($\Delta$ $\approx$ 10 cm$^{-1}$). A similar behavior is observed in other 2D materials such as the metal dichalcogenides MoS$_2$ \cite{Lee_2010} and TaSe$_2$ \cite{Castellanos_Gomez_2013} and has been associated with an enhancement of Coulomb screening as the number of layers decreases \cite{Molina_S_nchez_2011}. It is known that BSCCO-2212 undergoes a structural modulation affecting mostly the BiO planes along the crystalline $b$ direction, not only in the intrinsic compound but also in the doped one \cite{Levin_1994}. Micro X-ray diffraction in BSCCO-2212 exfoliated nanocrystals show that this modulation decreases significantly for crystals thinner than 60 nm \cite{Lupascu_2012}. The \textit{ab-initio} calculations of the Raman modes from \cite{Yelpo_2021} consider a correction of the crystalline structure due to the BiO planes modulation. Some of the Raman shifts found in this way are smaller compared to previous works where this distortion has been neglected \cite{Kovaleva_2004}. Therefore, it is possible to argue that the relaxation of the distortion as the layers becomes thinner can induce an increase in Raman frequency. Also, it is worth noting that the observed displacement of the Raman bands could be due to strain. Strain in the BSCCO flakes can be induced during the exfoliation process and also due to the interaction between the flake and the SiO$_2$/Si substrate \cite{Sahu_2019}. In addition, it is reasonable to think that the strain effect will be more prominent as the number of layers decreases.\\
As discussed previously with apparent color, Raman modes could also be affected by oxygen variations, and thus this could be the cause of the observed Raman-thickness shifts, as thin flakes are the most affected by air exposure \cite{Yu_2019, Zhao_2020}. We have studied the time evolution of the Raman modes at 116 cm$^{-1}$ and 460 cm$^{-1}$ for different flake thickness and no change in these bands can be seen.\\
The Raman shift – thickness relationship could be exploited as an additional method to characterize the thickness of exfoliated flakes together with the apparent color as discussed above. For both analyzed Raman modes, the obtained Raman frequency shift with thickness can be qualitatively fitted by the equation of the form \cite{Wang_2009}:
\begin{equation}
    \omega(n)=\omega_0+\frac{A}{n^B}
\end{equation}
where n is the number of UCs, $\omega_0$ is the bulk value of the Raman modes and A and B are fitting constants. The obtained parameters for the 116 cm$^{-1}$ band are: $\omega_0$=116 cm$^{-1}$, A=5.62, B=0.48 and for the 460 cm$^{-1}$ band: $\omega_0$=462 cm$^{-1}$, A=27.33, B=0.86. For thin enough flakes, the 460 cm$^{-1}$ band starts to overlap with the 520 cm$^{-1}$ Si Raman mode, making the application of formula (3) very difficult. 
\section*{IV.Conclusions}
We have mechanically exfoliated thin flakes of the high temperature superconductor BSCCO-2212 from a bulk single crystal. We have demonstrated that optical microscopy and micro-Raman spectroscopy are powerful tools to investigate the properties of this vdW material, allowing us to determine the flake thickness from different methods. By measuring apparent color, we have obtained a color-thickness chart of the crystal flakes. Moreover, by monochromatic illumination we have determined the refractive index of BSCCO-2212 flakes in the visible range, being the first work to report the refractive index of this material in the visible range by this technique. By modelling the BSCCO-2212/SiO$_2$/Si heterostructure using the Fresnel laws, we have calculated the best conditions in terms of SiO$_2$/Si substrates for the OC identification of flakes with different thickness.  We found that for 1 UC, illumination with 550 nm would rend the best OC values for easy identification. Finally, by micro-Raman spectroscopy we have characterized the Raman modes of this material and their evolution with flake thickness. A Raman hardening of the two most intense modes is identified as the flake thickness decreases. This hardening could be due to structural changes and strain induced during exfoliation and/or by the substrate interaction. A theoretical study, on the evolution of Raman modes, considering the thickness reduction, possible strain induced by exfoliation and the interaction with the substrate and oxygen content variation can contribute to the understanding on the origin of the observed Raman mode shifts in BSCCO-2212. Finally, we discussed the use of a phenomenological description for the evolution of the Raman bands positions that can be used for flake thickness determination. As BSCCO-2212 is a complicated material to investigate due to its sensitivity to ambient conditions deriving in an insulating state, the present work sheds light on BSCCO-2212 flakes optical and structural properties, paving the way for complex 2D vdW heterostructures fabrication for studying properties of unconventional superconductivity. .
\section*{Acknowledgments}
We thank A. Parente and G. Caballero for their help with the AFM measurements and W.Smith for the construction of the glove-box. I.F.C. also thanks E. Alcón for prolific discussions.  This work was supported by the Spanish Ministry for Science and Innovation (MCIN) under projects PGC2018-098613-B-C21 (SporQuMat) PID2021-122980OB-C52 (ECoSOx-ECLIPSE) and PID2021-124585NB-C33. I.F.C., E.M.G. and M. M. acknowledge support from the “Severo Ochoa” Programme for Centres of Excellence in R\&D (Grants SEV-2016-0686 and CEX2020-001039-S).  I.F.C holds a FPI fellowship from AEI-MCIN (PRE2020-092625).  A.S. acknowledges the financial support from MCIN for a Ramón y Cajal contract (No. RYC2021-031236-I), which is funded by the Recovery, Transformation and Resilience plan.
\section*{ORCID IDs}
\noindent
Ignacio Figueruelo-Campanero: \href{https://orcid.org/0000-0001-5144-9375}{0000-0001-5144-9375}\\
Adolfo del Campo: \href{https://orcid.org/0000-0002-9221-0587}{0000-0002-9221-0587}\\
Gladys Nieva: \href{https://orcid.org/0000-0001-9889-9875}{0000-0001-9889-9875}\\
Elvira M. Gonzalez: \href{https://orcid.org/0000-0001-9360-3596}{0000-0001-9360-3596}\\
Aida Serrano: \href{https://orcid.org/0000-0002-6162-0014}{0000-0002-6162-0014}\\
Mariela Menghini: \href{https://orcid.org/0000-0002-1744-798X}{0000-0002-1744-798X}\\
\bibliographystyle{unsrt}
\bibliography{references}
\end{multicols}

\end{document}